\documentclass[aps,prl,twocolumn,showpacs,superscriptaddress,floatfix]{revtex4}

\usepackage{amsmath}
\usepackage{amssymb}
\usepackage{amsbsy}
\usepackage{euscript}

\usepackage{graphicx}
\usepackage[usenames]{color}

\begin{document}

\title{Decoding sequential vs non-sequential two-photon double ionization of helium using nuclear recoil}

\author{D. A. Horner}
\affiliation{Los Alamos National Laboratory, Theoretical Division,
Los Alamos, NM 87545}
\author{T. N. Rescigno}
\affiliation{Lawrence Berkeley National Laboratory, Chemical Sciences,
Berkeley, CA 94720}
\author{C. W. McCurdy}
\affiliation{Lawrence Berkeley National Laboratory, Chemical Sciences,
Berkeley, CA 94720}
\affiliation{Departments of Applied Science and Chemistry, University of California, Davis,
CA 95616}

\date{\today}
\begin{abstract}
Above 54.4 eV, two-photon double ionization of helium is dominated by a sequential absorption process, producing characteristic behavior in the single and triple differential cross sections. We show that the signature of this process is visible in the nuclear recoil cross section, integrated over all energy sharings of the ejected electrons, even below the threshold for the sequential process. Since nuclear recoil momentum imaging does not require coincident photoelectron measurement, the predicted images present a viable target for future experiments with new short-pulse VUV and soft X-ray sources.

\end{abstract}


\maketitle

Kinematically complete experiments on one-photon double ionization of the simplest atomic~\cite{Schwarzkopf} and molecular~\cite{Weber} targets, coupled with state-of-the-art non-perturbative theoretical studies, have yielded fundamental information and insight into the nature of electron-electron correlation. With the advent of  new classes of short-pulse, high intensity free-electron laser~\cite{Ayvazyan} and high-harmonic  generation~\cite{Sekikawa} light sources that operate in the VUV and soft X-ray regimes, one looks forward to similar experiments involving the nonlinear process of few-photon multiple ionization~\cite{Benis,Moshammer}. With plans for such experiments currently underway, it is important to understand what kinds of new phenomena can be studied and the measurements that are most likely to be successful in revealing new effects~\cite{Lambropoulos}. 

Helium offers an interesting case in point. The energy required to doubly ionize helium is 79.0 eV, the sum of the first (24.6 eV) and second (54.4 eV) ionization energies. Double ionization of helium by two-photon absorption therefore requires a minimum photon energy of 39.5 eV. For photon energies between  39.5 and 54.4 eV the process is necessarily direct or nonsequential (NSI), ie. it requires simultaneous absorption of two photons
to strip two electrons from the atom and the process is expected to be sensitive to electron-electron correlation. For photon energies above 54.4 eV, sequential ionization (NSI) is possible: He$^+$ is produced by absorption of one photon, followed by absorption of  a second photon to produce He$^{++}$. Although NSI can compete with SI at these higher energies, SI, which is an essentially uncorrelated process, will dominate. In NSI, the excess energy ($2\hbar\omega - 79$~eV) can be shared continuously between the photoejected electrons, whereas with SI, we expect to see photoelectron energies sharply peaked around $\hbar\omega - 24.6$~eV and $\hbar\omega - 54.4$~eV. Experimental studies~\cite{Midorikawa_PRA05,Sorokin}, and indeed most theoretical treatments, of this process have to date focused on energy and angle-integrated quantities. In contrast to the one-photon, double ionization case, kinematically complete experiments that could distinguish between NSI and SI have yet to appear. Our purpose here is to demonstrate, using the results of precise quantum mechanical calculations, that since the angular distributions produced in two-photon NSI and SI  are very different, they leave a clear signature in the differential nuclear recoil distributions, even when integrated over all possible energy sharings of the ejected electrons. We will show that the signature of SI can be seen in the photoelectron energy-integrated nuclear recoil cross sections at energies several eV below the 54.4 eV threshold where sequential ionization is still a virtual process. Since these distributions do not require direct detection of the photo-emitted electrons, they present a attractive target for an experiment that could demonstrate a clear signature of sequential versus nonsequential double ionization. 

The triple differential cross section (TDCS) for two-photon double ionization is defined as
\begin{equation}
\label{eq:cross}
\frac{d\sigma}{dE_1d\Omega_1d\Omega_2}=\frac{2\pi}{\hbar}\frac{(2\pi\alpha)^2}{m^2\omega^2}k_1k_2
|f({\bf k}_1,{\bf k}_2,\omega)|^2 \, ,
\end{equation}
where $\mathbf{k}_1$ and $\mathbf{k}_2$ are the momenta of the photoelectrons, $\omega$ is the photon frequency, $m$ is the electron mass, $\alpha$ is the fine-structure constant, and $E_1=k_1^2/2$ is the energy of one of the electrons and thus defines the energy sharing.   The ionization amplitude   $f({\bf k}_1,{\bf k}_2,\omega)$ is in turn given by
\begin{equation}
\label{eq:amp}
f({\bf k}_1,{\bf k}_2,\omega)=\langle \Psi^-_{\mathbf{k}_1,\mathbf{k}_2}|\mu[E_0+\hbar\omega-H+i\epsilon]^{-1}\mu|\Phi_0\rangle \, .
\end{equation}
$H$ is the atomic Hamiltonian, $\Phi_0$ is the initial state of the atom with corresponding energy $E_0$,  and  $\Psi^-_{\mathbf{k}_1,\mathbf{k}_2}$ is the full momentum-normalized scattering wave function, with incoming boundary conditions corresponding to two free electrons,.  For polarization $\boldsymbol{\epsilon}$, the dipole operator in the velocity form,  $\mu$,  is defined in terms of the momentum operators, $\mathbf{p}_i$, for the two electrons by  $\mu =\boldsymbol{\epsilon}\cdot \mathbf{p}_1 + \boldsymbol{\epsilon}\cdot \mathbf{p}_2$.

The accompanying momentum recoil, $\mathbf{Q}$, of the nucleus due to the ejection of two electrons of momenta $\mathbf{k}_1$ and $\mathbf{k}_2$ is
\begin{equation}
\label{eq:Q}
\mathbf{Q}=-(\mathbf{k}_1+\mathbf{k}_2)
\end{equation}
At a given photon energy, we can therefore define a nuclear recoil cross section, differential in energy sharing and the angular dependence of $\mathbf{Q}$, by
\begin{equation}
\frac{d \sigma}{d^3 \mathbf{Q} dE_1} =  \int d\Omega_1 \int d\Omega_2  \, \frac{d \sigma}{d \Omega_1 d \Omega_2 dE_1} \, \delta^3(\mathbf{Q}+\mathbf{k}_1 + \mathbf{k}_2) 
\label{eq:nucrecoilxsec}
\end{equation}
The integral of this quantity over energy sharing, $d \sigma/d^3 \mathbf{Q} = \int dE_1 \, d \sigma/d^3 \mathbf{Q} dE_1$, carries a signature of the sequential ionization process even below its threshold, as we will see shortly.

We have recently shown~\cite{Horner2hv07} that task of calculating the two-photon, double ionization amplitude, which requires in principle the exact wave function for three-body Coulomb breakup, can be simplified by using the method of exterior complex scaling (ECS) which avoids the explicit imposition of asymptotic three-body boundary conditions~\cite{TopicalReview04}. 
We begin with the coupled driven equations in the Dalgarno-Lewis form of second order perturbation theory~\cite{DalgarnoLewis} that describe the absorption of two photons by a system initially in state $\Phi_0$,
\begin{equation}
\label{eq:Dalgarno1}
(E_0+\hbar\omega-H)\Psi^{\textrm{sc}}_1(\mathbf{r}_1,\mathbf{r}_2)=\mu\Phi_0
\end{equation}
\begin{equation}
\label{eq:Dalgarno2}
(E_0+2\hbar\omega-H)\Psi^{\textrm{sc}}_2(\mathbf{r}_1,\mathbf{r}_2)=\mu\Psi^{\textrm{sc}}_1 \,,
\end{equation}
which must be solved with pure outgoing wave boundary conditions for the wave functions $\Psi^{\textrm{sc}}_1$ and $\Psi^{\textrm{sc}}_2$.  With ECS, Eqs.~(\ref{eq:Dalgarno1}) and (\ref{eq:Dalgarno2}) are solved numerically on a discretized grid in a large but finite region of coordinate space extending to some  $R_0$, where the outer boundary conditions are obviated by rotating the radial coordinates beyond that point into the complex plane.  Having solved the driven equations, 
the ionization amplitude can then be extracted  using a  surface integral that involves a pair of testing functions $\psi^-_\mathbf {k}(\mathbf{r})$ which are momentum-normalized, one-electron Coulomb functions with nuclear charge $Z$=2, in the case of helium~\cite{McCurdy04}:
\begin{equation}
\begin{split}
f(\mathbf{k}_1,\mathbf{k}_2,\omega)
&= \frac{1}{2}  \int  \left(\psi^{-*}_\mathbf {k_1}(\mathbf{r_1}) \psi^{-*}_\mathbf {k_2} (\mathbf{r_2}) \, \nabla \,  \Psi^{\textrm{sc}}_2(\mathbf{r}_1,\mathbf{r}_2) \right. \\
& \left. \qquad  -\Psi^{\textrm{sc}}_2 (\mathbf{r}_1,\mathbf{r}_2)\, \nabla  \,\psi^{-*}_\mathbf {k_1}(\mathbf{r_1}) \psi^{-*}_\mathbf {k_2} (\mathbf{r_2}) \,  \right)  \cdot d\mathbf{S}
 \end{split}
\label{eq:3Damp}
\end{equation}
The integral is evaluated over a finite hypersphere whose radius is less than $R_0$.
No approximation concerning the final state has been made in this formalism because the testing functions, $\psi^-_\mathbf {k}(\mathbf{r})$ , merely extract the double ionization amplitude from  the final outgoing wavefunction  $\Psi_2^{\textrm{sc}} $, and electron correlation is treated completely in $\Psi_2^{\textrm{sc}}$ as well as in the initial state $\Phi_0$ in this approach.

But there is a technical difficulty must be addressed. For photon energies greater than the first ionization potential of the atom, $\Psi_1^{\textrm{sc}} $, the solution of Eq.~(\ref{eq:Dalgarno1}), will have undamped outgoing wave behavior on the real portion of the grid. Since $\Psi^{\textrm{sc}}_1$ serves as the source term in Eq. (\ref{eq:Dalgarno2}),  
$\Psi_2^{\textrm{sc}} $ will not converge with increasing $R_0$. As explained if ref.~\cite{Horner2hv07}, we can solve this problem by adding a small positive imaginary part to $\omega$ in Eq.~(\ref{eq:Dalgarno1}) only, rendering $\Psi_1^{\textrm{sc}} $ square-integrable, and then numerically extrapolating the results to real $\omega$.

$\Psi_1^{\textrm{sc}} $ and $\Psi_2^{\textrm{sc}} $ were expanded in a product basis of spherical harmonics, giving a set of coupled two-dimensional radial equations that were discretized using a finite element, discrete variable representation (DVR)~\cite{Rescigno00} and solved on parallel computers using sparse matrix methods. For the results reported here, we used partial waves up to $l=9$ and radial grids with real parts extending to 160 bohr on a side, with finite element boundaries starting at 5 bohr and then spaced 10 bohr apart, discretized with 18th order DVR. Calculations were performed for a range of complex $\omega$ values between  Im$(\omega)=0.500$ and Im$(\omega)=0.05$~hartrees and the individual partial wave amplitudes were then extrapolated to real $\omega$.

\begin{figure}
\begin{center}
\includegraphics[width=0.75\columnwidth,clip=true]{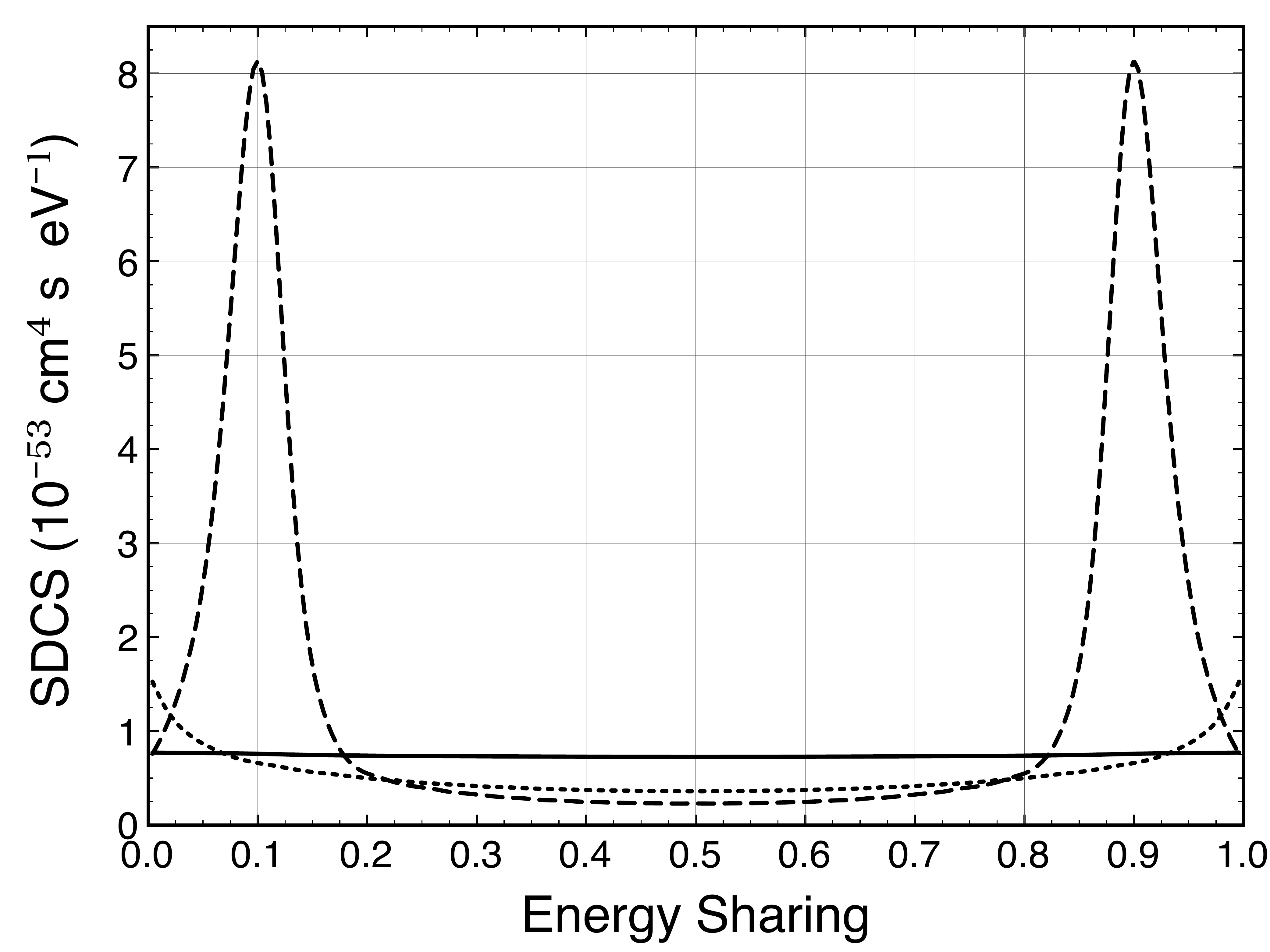}
\end{center}
\caption{Single differential cross sections at 44, 52 and 58 eV. }
\label{fig:sdcs}
\end{figure}

Figure~\ref{fig:sdcs} shows the single differential cross section (SDCS), which is obtained by integrating the  TDCS defined in Eq.~(\ref{eq:cross}) over the angles $\Omega_1$ and $\Omega_2$, at three different photon energies. The SDCS is seen to be a relatively flat function of energy sharing below 50~eV. The signature of SI becomes evident above 50~eV where the SDCS  begins to rise at the extremes of energy sharing until the sequential limit is reached at 54.4~eV. Above that energy, it has well defined peaks at E$_1=\hbar\omega-$54.4~eV and E$_1=\hbar\omega-$24.6~eV. Our calculated SDCS at 58~eV has finite peak heights because it was extrapolated from calculations carried out on a finite grid. These peaks would become singularities in the limit of an infinite grid, and are a fundamental feature of the cross sections derived in lowest order perturbation theory.  The apparent widths of the peaks, however, are not a consequence of the finite grid, as explained below.

\begin{figure}
\begin{center}
\includegraphics[width=0.75\columnwidth,clip=true]{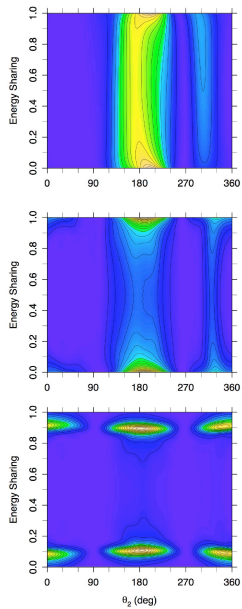}
\caption{Triple differential cross sections, as a function of energy sharing and angle, at (top to bottom) 44, 52 and 58 eV, showing the signatures of nonsequential and sequential ionization (see text).  The direction of one electron is fixed at 30$^{\circ}$ to the polarization direction while the other is varied.}
\label{fig:tdcs}
\end{center}
\end{figure}

One also sees a striking change in the TDCS as we move from regions dominated by NSI processes to those where SI dominates. Figure~\ref{fig:tdcs} shows contour plots of the TDCS, as a function of energy sharing and scattering angle for one of the ejected electrons, at photon enertgies of 44, 52 and 58~eV. The scattering angle of the second electron is fixed at 30$^\circ$. At 44~eV, the TDCS has its maximum value when the two electron are ejected back-to-back, independent of the energy sharing, which is expected for a correlated, NSI process. At 52~eV, the angular dependence is more complicated, but one still sees a propensity for back-to-back ejection, as well as a peak near 180$^\circ$ for extreme unequal energy sharing. At 58~eV, the TDCS is strongly peaked near 0 and 180$^\circ$ at the energy sharings corresponding to SI and is uniformly small elsewhere.
\begin{figure}
\begin{center}
\includegraphics[width=0.75\columnwidth,clip=true]{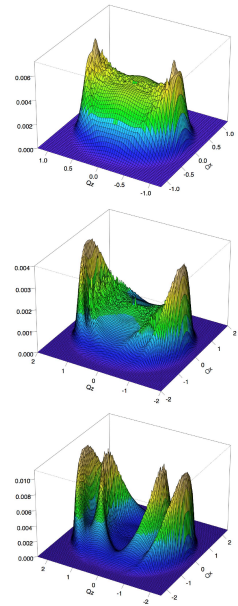}
\caption{Nuclear recoil cross sections, integrated over all photoelectron energy sharings, at 44, 52 and 58 eV. The polarization vector is chosen to lie along the z-axis.}
\label{fig:recoil}
\end{center}
\end{figure} 

As we showed earlier~\cite{Horner2hv07}, we can get a qualitatively correct description of sequential ionization with a simple analytic model that ignores correlation in the initial, intermediate and final states. That model, whose details are derived elsewhere~\cite{nucrec}, gives the following expression for the TDCS,
\begin{equation}
\begin{split}
& \frac{d\sigma^{\textrm{seq}}}{d E_1 d\Omega_1 d\Omega_2} \approx 
\frac{\hbar}{4 \pi} 
\left(\frac{3}{4\pi}\right)^2 \cos^2(\theta_1) \cos^2(\theta_2)  \\
&\left[\frac{\sqrt{\sigma^{\rm{He}^+}(E_2)  \sigma^{\rm{He}}(E_1)}}{E_0+\hbar \omega -\epsilon_{1s}-E_1} 
 + \,\,\frac{\sqrt{\sigma^{\rm{He}^+}(E_1) \sigma^{\rm{He}}(E_2)}}{E_0+\hbar \omega -\epsilon_{1s}-E_2} \right] ^2\, ,
\end{split}
\label{eq:seqapproxtdcs}
\end{equation}
where $\sigma^{\rm{He}}$ and $\sigma^{\rm{He}^+}$ are the photoionization cross sections of helium and He$^+$, respectively, and $\epsilon_{1s}$ is the energy of He$^+$.  When integrated over $\Omega_1$ and $\Omega_2$, Eq.(\ref{eq:seqapproxtdcs}) gives an SDCS proportional to the square of the term in square brackets.  Thus the model  shows that when the photon energy exceeds 54.4 eV ($\epsilon_{1s}$) the TDCS and SDCS diverge at the ejected electron energies corresponding to SI, the singularities  separated by the difference between the ionization energies of He and He$^+$.  The model also explains the rapid rise of the SDCS just below the SI threshold at extreme unequal sharing, since  an apparent  width (determined by the size of the numerators in Eq.(\ref{eq:seqapproxtdcs})) is associated with the singularities above the SI threshold which causes the SDCS to rise above the NSI background  .   As mentioned above, these singularities are consequences of treating sequential ionization in lowest order perturbation theory, and not of the simplifying assumptions used in deriving the model.

The simple model also shows that the angular dependence of the TDCS is the product of two uncorrelated dipole distributions for each ejected electron. This behavior is clearly seen in the calculated TDCS at 58~eV which shows peaks at $\theta_2=0^\circ$ and 180$^\circ$. The calculated TDCS can be used to compute the momentum recoil imparted to the nucleus.

Figure~\ref{fig:recoil} shows the calculated nuclear recoil cross sections, integrated over all energy sharings, at the three photon energies previously considered. The integration in Eq.(\ref{eq:nucrecoilxsec})  requires some effort  because, for example,  there are ranges directions of $\mathbf{k}_1$, for which no $\mathbf{k}_2$ exists that satisfies Eq.(\ref{eq:Q}). Details about the numerical evaluation of $d \sigma/d^3 \mathbf{Q} dE_1$ will be given elsewhere~\cite{nucrec}.  The general shapes of these cross sections can be understood on the basis of the previously described TDCS. For NSI, the two electrons are preferentially ejected back-to-back, which means comparatively little momentum is imparted to the nucleus. But the imparted momentum for back-to-back ejection is exactly zero only for equal energy sharing and we are integrating over all energy sharings, so we would expect a roughly isotropic distribution peaked about $\mathbf{Q}=0$ for a pure NSI process. For SI, however, the electrons are uncorrelated and, because of their $\cos^2(\theta)$ angular dependence, are  preferentially ejected along the polarization axis, i.e.  at 0$^\circ$ or 180$^\circ$. We therefore expect, for a pure SI process, that the distribution of nuclear recoil momenta will show four peaks along the polarization axis at $|\mathbf{Q}|=\pm(|\mathbf{k_1}| + |\mathbf{k_2}|)$ and $|\mathbf{Q}|=\pm(|\mathbf{k_1}| - |\mathbf{k_2}|)$. These general features are indeed seen in the calculated cross sections. The nuclear recoil cross sections at 58~eV show four clearly defined rings (whose peak heights we must emphasize would diverge if we could use an infinite grid). Even at 52~eV, we can see a clear signature of SI in the nuclear recoil cross section where there are two distinct rings at the extremes of momentum transfer along with two secondary wings  developing inside the  prominent wings -- the signature of ``virtual SI''. 

In summary, we have shown that  sequential ionization leaves a clear signature in the nuclear recoil cross section, even when it is integrated over all energy sharings of the ejected electrons. This signature is clearly visible at 52~eV, which is 2.4~eV below the SI threshold. The nuclear recoil cross section is found to be a very sensitive indicator of SI. We see "virtual SI" structure in the nuclear recoil cross section even at 44~eV, while the SDCS at the same photon energy is flat. The energy-integrated nuclear recoil can be measured in a double ionization experiment using, for example, the COLTRIMS technique~\cite{Ullrich}. Such observations do not require detection of the electrons at all and thus avoid completely the need for coincidence measurements.

This work was performed under the auspices of the US Department of Energy
by the Los Alamos National Laboratory and the University of California Lawrence Berkeley National Laboratory
under Contract DE-AC02-05CH11231 and
was supported by the U.S. DOE Office of Basic Energy
Sciences, Division of Chemical Sciences. CWM acknowledges support from the National Science Foundation (Grant No. PHY-0604628).

%


\end{document}